\documentclass[lettersize,journal]{IEEEtran}
\usepackage{amsmath,amsfonts}
\usepackage{algorithmic}
\usepackage{array}
\usepackage[caption=false,font=normalsize,labelfont=sf,textfont=sf]{subfig}
\usepackage{textcomp}
\usepackage{stfloats}
\usepackage{url}
\usepackage{verbatim}
\usepackage{graphicx}
\def\BibTeX{{\rm B\kern-.05em{\sc i\kern-.025em b}\kern-.08em
    T\kern-.1667em\lower.7ex\hbox{E}\kern-.125emX}}
\usepackage{balance}

\usepackage{bm}

\usepackage{amsmath}

\usepackage{hyperref}

\newcommand{\norm}[1]{\left\lVert #1 \right\rVert}

\begin{document}
\title{Lexicographic optimization for real-time CNC feedrate planning with coupled orientation handling}

\author{Haijia Xu and Alexander Verl
\thanks{Haijia Xu and Alexander Verl are with the Institute for Control Engineering of Machine Tools and Manufacturing Units, University of Stuttgart, Stuttgart 70174, Germany (e-mail: haijia.xu@isw.uni-stuttgart.de; alexander.verl@isw.uni-stuttgart.de).}
}

\markboth{Journal of \LaTeX\ Class Files,~Vol.~18, No.~9, June~2026}%
{How to Use the IEEEtran \LaTeX \ Templates}

\maketitle

\begin{abstract}
Optimization-based feedrate planning offers the potential to significantly increase machining productivity, but its industrial adoption has been limited by high computational cost and extensive tuning effort. 
This paper proposes a lexicographic feedrate optimization principle that adaptively balances finishing time and motion smoothness in a tuning-free manner. 
To further improve computational efficiency, the optimization scheme is extended by a sparsity-exploiting formulation combined with a sequential windowing strategy, enabling real-time capable execution. 
In addition, a unified toolpath parameterization scheme is incorporated to synchronously handle tool position and orientation within the optimization framework. 
For a five-axis freeform test contour, the proposed method takes 14~s on an Intel i5-3470 CPU to optimize feedrate profiles for long toolpaths with 100\,000 constraint checkpoints, and 52~s on a high-performance AMD 9950X CPU to handle one million checkpoints. Compared to an industrial CNC kernel, the resulting finishing time is reduced by more than 15~\%.
\end{abstract}


\section{Introduction}
\begin{figure}[htb]
  \centering
  \includegraphics[width=0.45\textwidth]{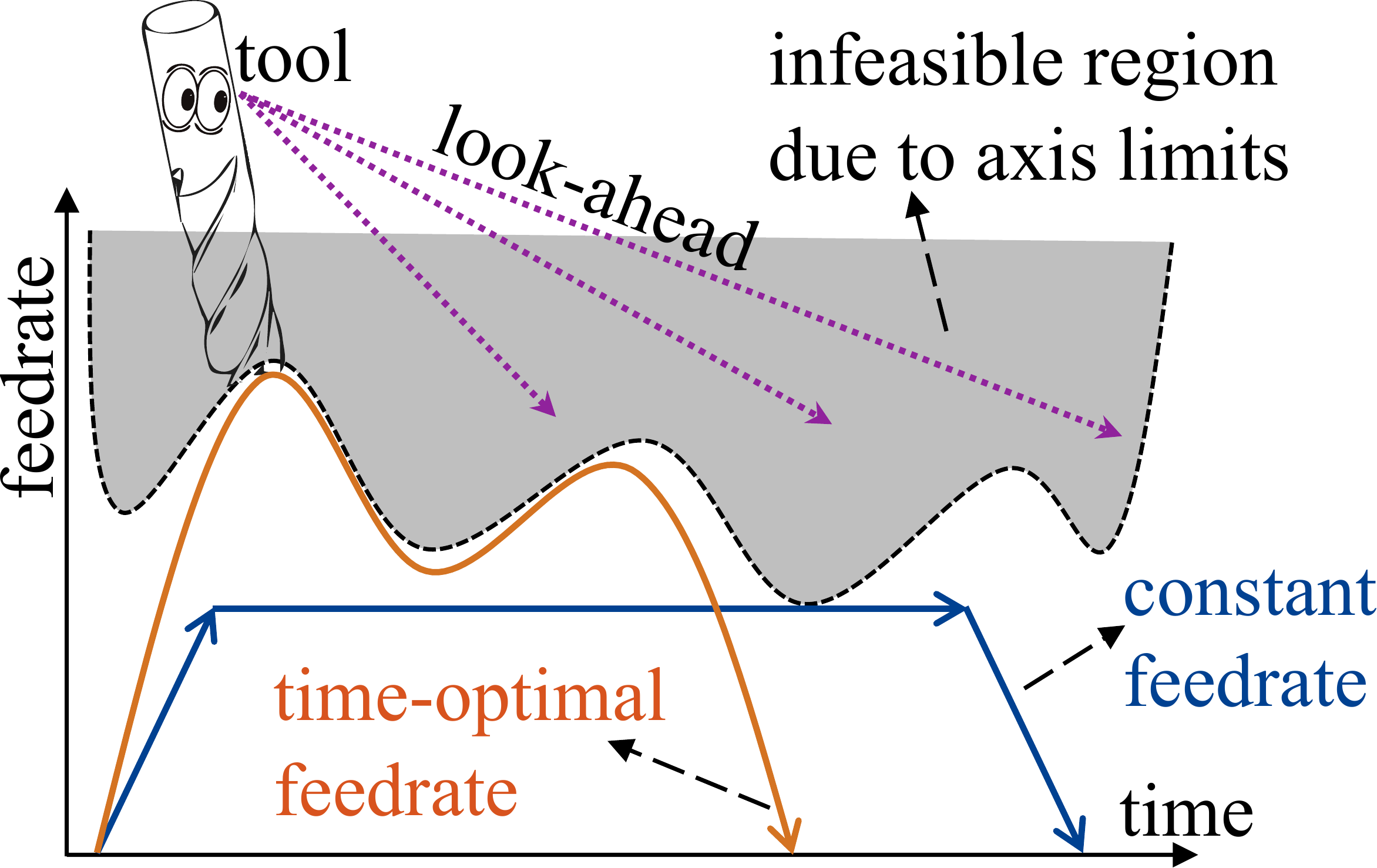}
  \caption{Illustration of constant and minimum-time feedrate profiles. } 
  \label{fig:LookAhead}
\end{figure}

Manufacturing technology increasingly demand for higher productivity by fully exploiting the dynamic capability of servo drives, such as velocity and acceleration limits, to maximize the feedrate while maintaining the machining accuracy. 
However, due to the strict requirement of real-time capability, motion planning algorithms implemented in industrial CNC kernels still rely on profile-based methods \cite[§7.2.1]{Altintas2011}.
These methods repeatedly fit feedrate profiles on the Cartesian toolpath and transform them back into axis space, 
relying on iterative or approximate procedures to satisfy actuator constraints.
Moreover, the use of predefined feedrate templates, such as seven-phase profile \cite{Erkorkmaz2001}, 
imposes a fixed trajectory structure with piecewise linear characteristics (e.g., bounded feedrate and acceleration segments).
Such parameterizations are unable to accurately capture the curvature-dependent nature of the time-optimal feedrate profile, which generally exhibit freeform shapes determined by the toolpath geometry and actuator constraints. Consequently, this heuristic planning approach inherently compromises time optimality, as illustrated in \autoref{fig:LookAhead}.

Various optimization-based algorithms have been explored in recent literature to compute time-optimal feedrate profiles under actuator limits. These include second-order cone programming \cite{Verscheure2009}, sequential quadratic programming \cite{Sencer2008}, and interior point method \cite{DiCola2025}, to name a few. Despite their improved time optimality, these approaches have not found their way into industrial CNC kernels due to their significantly increased computational burden. The resulting calculation times are far too long for real-time CNC feedrate planning, particularly for long toolpaths.

This paper aims to further narrow the gap between the theoretical development of feedrate optimization algorithms and their practical CNC applications. To this end, we first present an optimization framework capable of real-time execution that builds on the strengths of linear programming (LP) formulations \cite{Fan2013,Erkorkmaz2017}. By exploiting the sparse problem structure and integrating the solver closely with the CNC look-ahead functionality, the method achieves verified real-time performance on a single CPU core, even for long toolpaths and on legacy hardware. In addition, we introduce a lexicographic design principle that explicitly captures the multi-objective nature of feedrate planning by minimizing finishing time while maximizing motion smoothness. This formulation optimizes the primary objective (finishing time) first and then the secondary objective (smoothness), while preserving the desired optimality of the first objective. This is done in an adaptive and tuning-free manner. Moreover, we demonstrate that the proposed framework can naturally be extended to handle tool orientation constraints in a coupled manner without additional synchronization. Experimental results on a five-axis machine tool illustrate the method's effectiveness and potential for integration into future CNC kernels. 

\begin{figure*}[htb]
	\centering
	\includegraphics[width=0.95\textwidth]{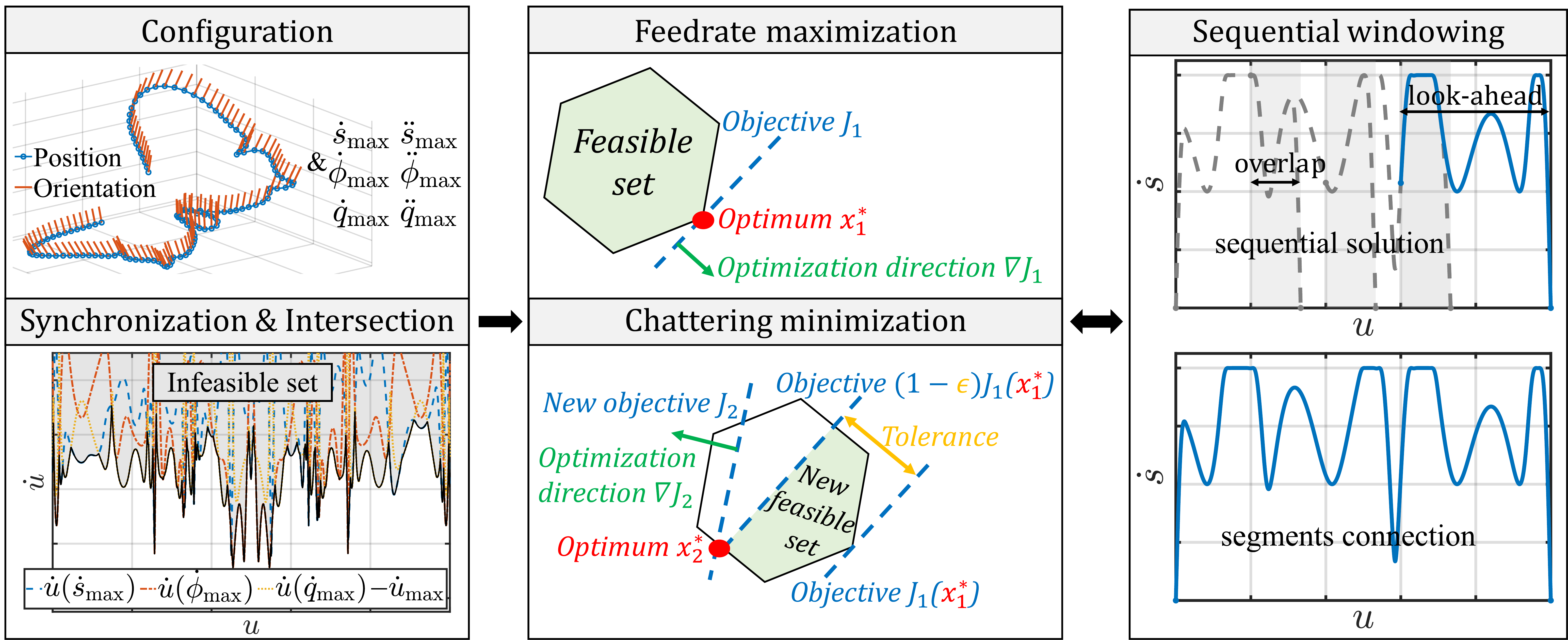}
	\caption{Proposed lexicographic LP framework for real-time, multi-objective feedrate optimization with synchronized position-orientation motion.} 
	\label{fig:generalProblem_and_solution}
\end{figure*}

The main contributions are twofold. First, a lexicographic design that enables adaptive feedrate smoothing without manual tuning.
Second, a sparse formulation with sequential windows whose efficiency has been verified for long toolpaths with up to one million checkpoints.
The key concept is shown in \autoref{fig:generalProblem_and_solution}.

\section{Lexicographic feedrate optimization}

This paper focuses only on the acceleration-limited planning. An extension to jerk constraints within the LP framework using the concept of pseudo jerk can be found in \cite{Fan2013}.

The following formulation considers the predefined toolpath $\bm r=[X,Y,Z,U,V,W]^T$ of a five-axis machine tool (with rotary b and c axes) in the workpiece coordinate system (WCS). Here, $\bm p=[X,Y,Z]^T$ and $\bm o=[U,V,W]^T$ represent the tool position and the tool orientation on the unit sphere, respectively.  Assuming that the second-order parametric derivatives of the toolpath data exist (e.g., by spline fitting \cite{Brecher2006}) and that the related inverse kinematics are available (see e.g., \cite{Jiang2020}), the toolpath is also considered to be given in the machine coordinate system (MCS) for each axis as:
\begin{align}
    \bm q(u) = \left[x(u), y(u), z(u), b(u), c(u) \right]^T,
	\label{eq:path}
\end{align}
with the normalized path parameter $u \in [0,1]$. In contrast to the decoupled approach \cite{Jiang2020}, this parametrization represents the axis translational and rotational motion in a geometrically coupled manner and inherently ensures motion synchronization. In addition, the toolpath geometry does not have to be given in arc length $s$ or arc radian $\phi$ for Cartesian constraints handling in WCS, respectively.
The axis level limits are given by
\begin{subequations}
    \begin{align}
        -\dot{\bm q}_\text{max}
        \leq \hspace{0.2em} \dot{\bm q}
        &= \bm q'(u) \dot{u} 
        \leq \dot{\bm q}_\text{max} \\
        -\ddot{\bm q}_\text{max} 
        \leq \hspace{0.2em} \ddot{\bm q}
        &= \bm q''(u) \dot{u}^2 + \bm q'(u) \ddot{u} 
        \leq \ddot{\bm q}_\text{max}.
    \end{align}
\label{eq:constrAxis}\end{subequations}
Here, $\bm q, \bm q''$ denote the geometric derivatives of the path $\bm q(u)$ with respect to the path parameter $u$, and $\dot{u}$, $\ddot{u}$ denote its time derivatives. The time derivative of the toolpath $\bm r$ is given by:
\begin{align}
    \dot{\bm r} = \bm J(\bm q(u)) \bm q'(u)\dot{u},
\end{align}
with the kinematic Jacobian matrix $\bm J = \partial \bm H / \partial \bm q \in \mathbb{R}^{6\times 5}$ derived from the forward transformation $\bm r = \bm H(\bm q)$. The tool tip feedrate value and its time derivative are limited by
\begin{subequations} 
\begin{align}
    0 \leq
    \dot{s}
    &= \psi_{\bm p}(u)\dot{u}  
    \leq \dot{s}_\text{max} \\
    -\ddot{s}_\text{max} \leq
    \ddot{s}
    &= \psi_{\bm p}'(u)\dot{u}^2 + \psi_{\bm p}(u) \ddot{u}
    \leq \ddot{s}_\text{max}.
\end{align}
\label{eq:constrToolPos}\end{subequations}
where the translational velocity transformation $\psi_{\bm p}(u) \in \mathbb{R}$ is given by the Euclidean norm $\psi_{\bm p}(u) := \norm{\left[\bm I, \bm 0\right] \bm J(\bm q(u)) \bm q'(u)}_2$, which maps the path parameter velocity  $\dot{u}$ to the feedrate $\dot{s}$.
Similarly, the constraints on the angular velocity and acceleration of the tool orientation are expressed as 
\begin{subequations} 
\begin{align}
    0 \leq
    \dot{\phi}
    &= \psi_{\bm o}(u)\dot{u}  
    \leq \dot{\phi}_\text{max} \\
    -\ddot{\phi}_\text{max} \leq
    \ddot{\phi}
    &= \psi_{\bm o}'(u)\dot{u}^2 + \psi_{\bm o}(u) \ddot{u}
    \leq \ddot{\phi}_\text{max},
\end{align}
\label{eq:constrToolOri}\end{subequations}
with the angular velocity transformation $\psi_{\bm o}(u) := \norm{\left[\bm 0, \bm I\right] \bm J(\bm q(u)) \bm q'(u)}_2 \in \mathbb{R}$.
Notably, for a given machine kinematics,  the velocity transformations $\psi_{\bm p}(u)$ and $\psi_{\bm o}(u)$ only depend on the toolpath $\bm q(u)$. 
Therefore, The parametric derivatives $\psi_{\bm p}'(u)$ and $\psi_{\bm o}'(u)$ can be derived analytically using the chain rule.

By defining a transformed optimization variable $b(u) = \dot{u}^2$, the constrained feedrate maximization problem can be equivalently written in a linear form for $u \in [0,1]$ as 
\begin{align}
    & \hspace{5em}  \max_{b(u)} \int_{0}^{1} b(u) \mathrm{d}{u} &  
    \label{method:linearFormulation}
    \\
    &\begin{aligned}
    %
    \text{s.t.} \quad 0 
    & \leq   \bm q'^2(u) b(u) 
    \leq \dot{\bm q}^2_\text{max} \\
    -\ddot{\bm q}_\text{max} &\leq 
    \bm q''(u) b(u) + 0.5 \bm q'(u) b'(u) 
    \leq \ddot{\bm q}_\text{max}  
     \nonumber
    \end{aligned}
    \\
    &\begin{aligned}
    %
    %
    0  & \leq  \psi^2_{\bm p}(u)b(u) 
    \leq \dot{s}^2_\text{max} \\
    -\ddot{s}_\text{max} &\leq 
    \psi_{\bm p}'(u) b(u) + 0.5 \psi_{\bm p}(u) b'(u)
    \leq \ddot{s}_\text{max}\\
    %
    %
    0  & \leq  \psi^2_{\bm o}(u) b(u) 
    \leq \dot{\phi}^2_\text{max} \\
    -\ddot{\phi}_\text{max} &\leq 
    \psi_{\bm o}'(u) b(u) + 0.5 \psi_{\bm o}(u) b'(u)
    \leq \ddot{\phi}_\text{max}
    \nonumber
    \end{aligned}
\end{align}

Numerical optimization algorithms typically terminate with a small residual error. This may lead to chattering in the optimized feedrate profiles, particularly when coarse discretization is required for feedrate optimization in the real-time environment. While such solutions still satisfy the given constraints, excessive in-bound chattering can severely degrade motion smoothness and thus the finishing quality. The need for improving smoothness can be given by the objective function
\begin{align}
    \min_{b(u)} \int_{0}^{1} \left| b''(u) \right| \mathrm{d}{u}.
	\label{obj:smoothing}
\end{align}
This objective penalizes rapid variations in the feedrate profile, improving motion smoothness in addition to limit satisfaction. However, from an optimization perspective, maximizing the feedrate as in \eqref{method:linearFormulation} while minimizing the chattering as in \eqref{obj:smoothing} are inherently competing objectives. A common strategy is to combine them into a single weighted objective function \cite{Xu2025IJMTM}, given by
\begin{align}
    \min_{b(u)} \int_{0}^{1} - b(u) + \gamma \left| b''(u) \right| \mathrm{d}{u}.
	\label{obj:blend}
\end{align}
The weighting factor $\gamma$ enables a trade-off between time optimality and motion smoothness. Despite its simplicity, the choice of $\gamma$ is highly geometry-dependent. This requires case-specific tuning, which can be non-intuitive for long freeform toolpaths.

To address this issue, the principle of lexicographic optimization can be adopted to adaptively handle this trade-off in a tuning free manner, even under varying toolpath geometries. 
The basic idea is to first optimize the objective \eqref{method:linearFormulation} to obtain its optimum $b^{*\text{\eqref{method:linearFormulation}}}(u)$, and then optimize the objective \eqref{obj:smoothing} subject to an additional constraint on the tolerable degradation of the first objective, given by   
\begin{align}
    \int_{0}^{1} b(u) \, \mathrm{d} u \geq (1-\epsilon) \int_{0}^{1} b^{*\text{\eqref{method:linearFormulation}}}(u) \mathrm{d}u.
\end{align}
This constraint, with a factor $\epsilon>0$, specifies the tolerated relative degradation of the maximized feedrate profile while optimizing the smoothness objective, as illustrated in \autoref{fig:generalProblem_and_solution}. 

\begin{figure}[htb]
  \centering
  \includegraphics[width=0.475\textwidth]{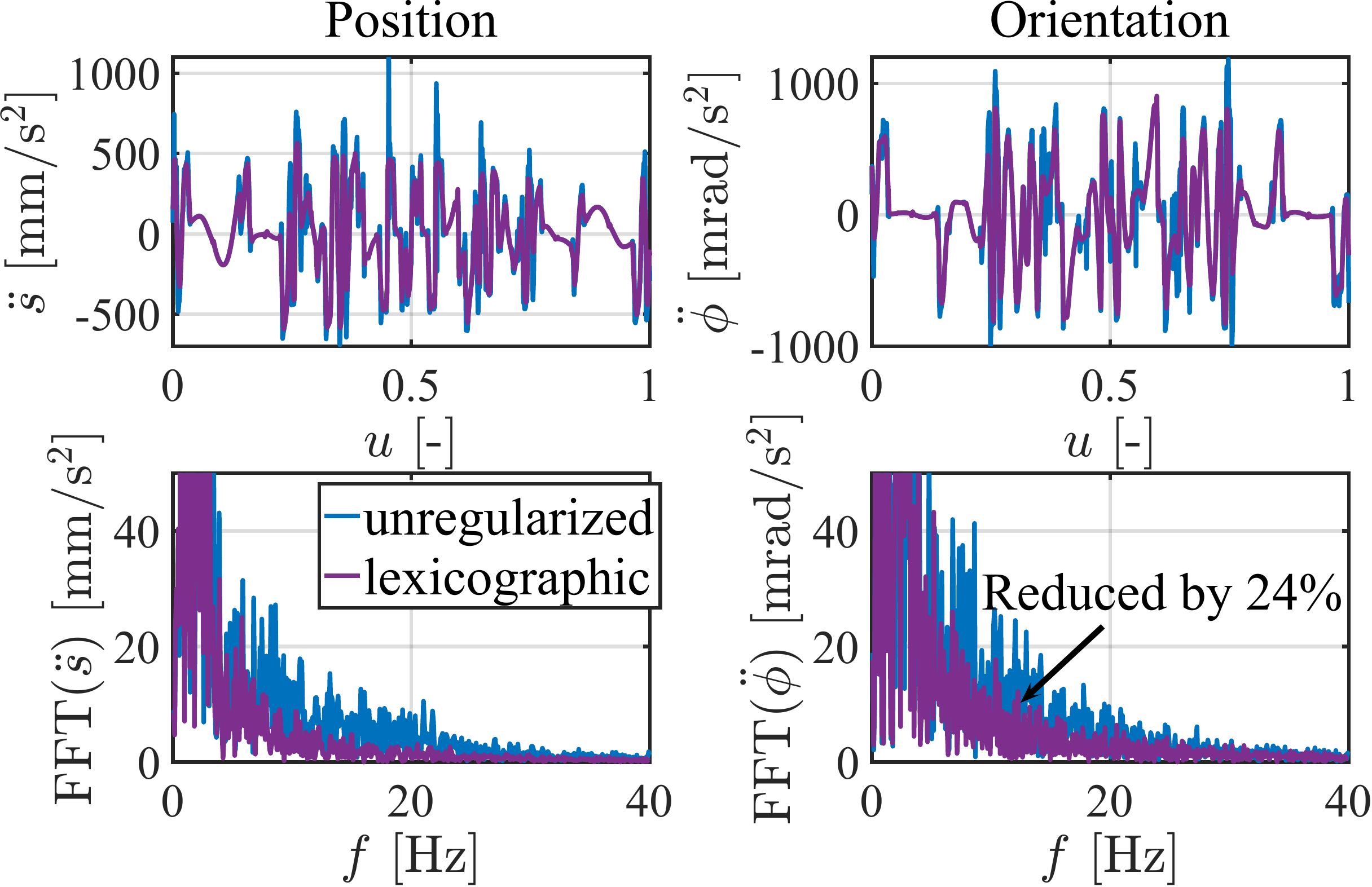}
  \caption{Smoothing effect of the proposed lexicographic optimization.} 
  \label{fig:LexSmooth}
\end{figure}

In the freeform test case for a five-axis machine tool (cf. \autoref{sec:validation}), setting the relative tolerance to $\epsilon=1\%$ reduces the acceleration-level chattering by $24$~\% without additional user intervention, at the cost of a $1.3$~\% increase in finishing time, as shown in \autoref{fig:LexSmooth}. The lexicographic design eliminates the need for case-specific tuning of the weight $\gamma$ in \eqref{obj:blend}, which lacks a clear physical interpretation. 
As a result, the optimization trade-off is handled adaptively in a manner that more closely aligns with engineering intuition.

\section{Sparse discretization and sequential windowing}
\begin{figure}[htb]
  \centering
  \includegraphics[width=0.25\textwidth]{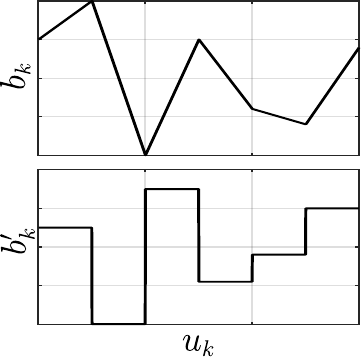}
  \caption{Discretized piecewise representation for the transformed feedrate $b(u_k)$.} 
  \label{fig:discretization}
\end{figure}

To obtain an efficiently solvable LP, the continuous formulation is discretized with a piecewise representation.
The path parameter $u \in [0,1]$ is evenly discretized with spacing $\Delta u = 1/N$, resulting in discretization grids $u_k = k \Delta u$ for $k=0,...,N$. The transformed variable $b(u)$ is represented by 
\begin{align}
    b_{k+1} = b_k + b'_k\Delta u,
	\label{eq:integrator}
\end{align}
where the second-order parametric derivative $b''_k$ used for motion smoothing is approximated by $b''_k = (b'_{k+1} - b'_{k})/ \Delta u$ to further reduce the optimization dimension. 
The discretized objective functions can be written as 
\begin{align}
    \max_{b_k} \sum_{0}^{N} b_k,
\end{align}
and
\begin{align}
     \min_{s_k} \quad  & \sum_{k=0}^{N} s_k \\
     \text{s.t.} \quad 
     -s_k &\leq \frac{b'_{k+1} - b'_k}{\Delta u} \leq s_k, \\
     s_k &\geq 0.
\end{align}
The discretized piecewise representation is illustrated in \autoref{fig:discretization}.
Notably, the nonlinear absolute value operator in \eqref{obj:smoothing} is equivalently reformulated in a linear form by introducing magnitude-bounded slack variables $s_k$. 
The equality constraint \eqref{eq:integrator} is checked at each discretization grid $k$ to enforce the integration rule. 
This approach reformulates the continuous integrator with respect to $u$, and correspondingly, the optimization problem, into a large sparse structure. 
Since most entries in this representation are zero, specialized numerical solvers, such as HiGHS \cite{Huangfu2018}, can be adopted to exploit the sparsity pattern to avoid unnecessary computations.

Despite the efficiency of LP, one-shot feedrate optimization for long toolpaths still remains impractical due to the significantly increased computational and memory demands. To ensure real-time capability, the numerical optimization is performed sequentially on overlapping windows, as shown in \autoref{fig:generalProblem_and_solution}. 
In contrast to the parallel windowing strategy in \cite{Erkorkmaz2017}, the presented sequential windowing approach does not require additional memory buffers or multiple CPU cores, making it compatible with legacy hardware commonly used in CNC systems.
In addition, the sequential processing procedure closely aligns with the CNC look-ahead functionality \cite{Zhao2013}, which provides geometric information for a short segment of the toolpath in the future.

The initial conditions of each subsequent window are taken from the solution of the previous window at the start of the overlap region. This ensures a consistent transient phase without the need for additional transient blending operations. The terminal velocity of each window is set to zero for safety in the event of solver failure. 
The choice of window length and overlap balances time optimality and computational load. A larger window length reduces the loss of time optimality in general, and should be selected based on the available computational resources. The overlap must be sufficiently long to allow to deceleration from maximum feedrate within the overlap region.

\begin{figure}[htb]
  \centering
  \includegraphics[width=0.475\textwidth]{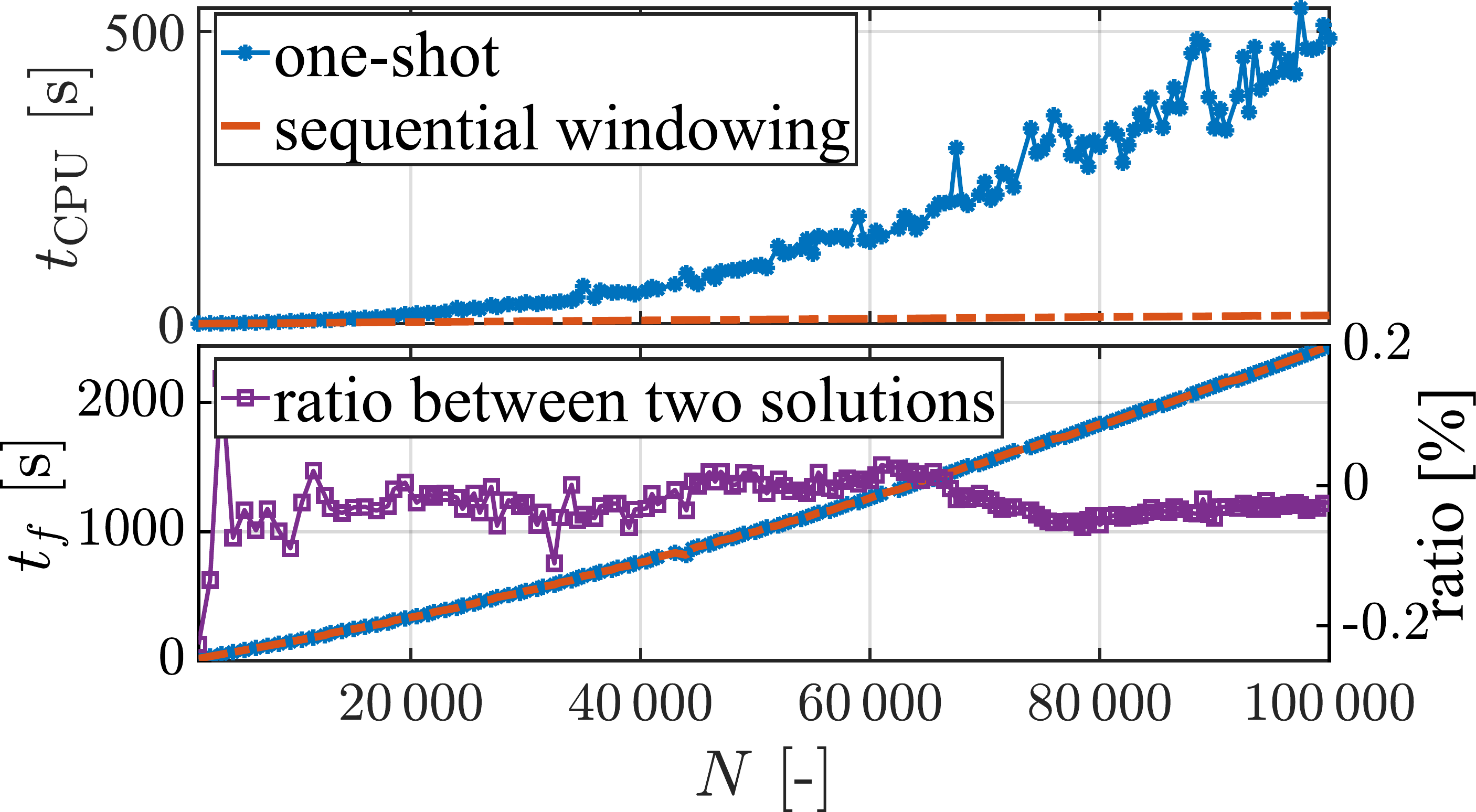}
  \caption{Comparison between one-shot and windowed solutions.} 
  \label{fig:Runtime}
\end{figure}

\autoref{fig:Runtime} shows the measured computational performance of the proposed sequential windowing (SeWin) strategy on legacy hardware with an Intel i5-3470 CPU (@3.60GHz) under single-core execution. The tests are conducted with a window length of 500 and an overlap of 200. The main results are:
\begin{enumerate}
	\item 	As opposite to the one-shot solution, which has a polynomial computational load, the proposed SeWin shows linear scalability with respect to the number of constraint checkpoints, and, correspondingly, the toolpath length. For example, processing $10\,500$ points takes $1.5$~s, while $100\,000$ points are processed in $14.3$~s. In addition, outliers of computational time occur frequently in the one-shot solution, especially for large dimensions. 
	\item 	The SeWin solution does not necessarily compromise the finishing time $t_f$ compared to the one-shot solution, as the relative time difference remains within $\pm 0.2$~\%. This effect becomes more pronounced for large-scale problems. Due to premature termination of the numerical solver, the one-shot solution does not always reach the global optimum.
	\item 	In an extreme case with one million constraint checkpoints, the proposed method took $t_\text{CPU} = 51.5$~s on a high-performance AMD 9950X CPU (@5.76GHz) and $t_\text{CPU} = 145.7$~s on a slower Intel i5-3470 CPU (@3.60GHz) to process an ultra-long toolpath with a finishing time $t_f$ exceeding $5.4$~h. These results further demonstrate the computational robustness of the presented optimization framework under extreme conditions and highlight its potential for integration into future CNC kernels for real-time feedrate optimization.
	\item 	Due to its high efficiency, fine discretization grids can be employed at the feedrate optimization stage to minimize the real-time interpolation errors in ultra-high-precision processes, such as laser processes. 
\end{enumerate}
Overall, combining the LP-based feedrate optimization algorithm with the sequential windowing strategy enables real-time capable computations for long toolpaths.
The computational efficiency and robustness have been
verified under single-core execution with up to one million constraint checkpoints.

\section{Validation in simulation and experiments}
\label{sec:validation}
The proposed lexicographic optimization method (LexLP) combined with the SeWin strategy is benchmarked against an industrial CNC kernel, which represents the only comparable approach that guarantee both constraint feasibility and real-time capability. In addition, the one-shot feedrate maximization solution without feedrate smoothing in \eqref{method:linearFormulation} is included as a baseline for comparison. 

\begin{figure}[ht]
  \centering
  \includegraphics[width=0.475\textwidth]{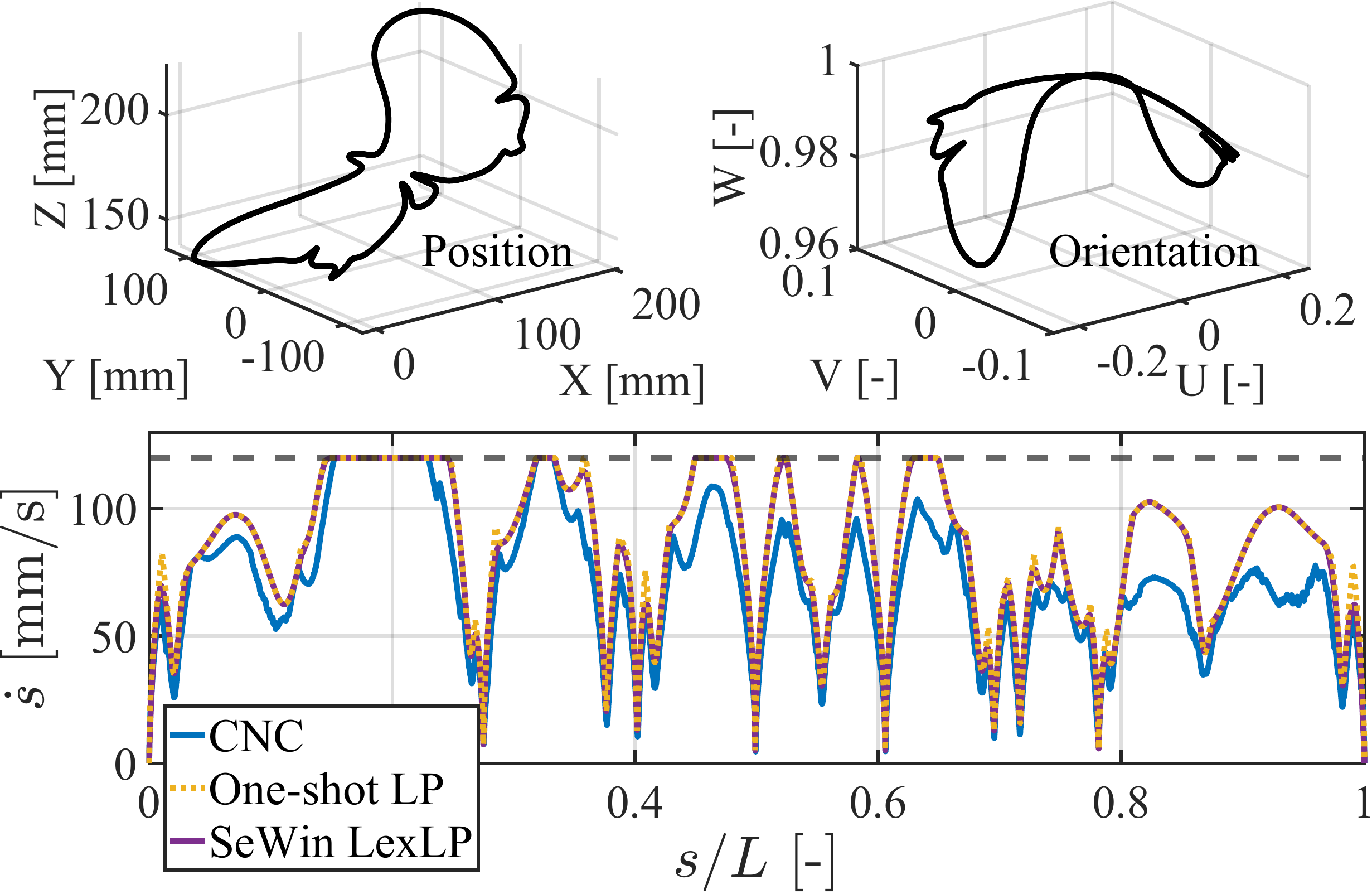}
  \caption{Feedrate comparison on a freeform toolpath with synchronized tool position and orientation.} 
  \label{fig:CompareFeed}
\end{figure}

The freeform test contour in \autoref{fig:CompareFeed} is used for the benchmark. 
This test contour is converted into G01 linear segments for execution on the TwinCAT 3 CNC kernel, which are smoothed using the High Speed Cutting (HSC) surface optimizer \cite{HSCsurface} with an allowed path deviation of $0.2$~mm. The axis limits are set to $\dot{q}_\text{max} = 80$~mm/s and $400$~mrad/s, and $\ddot{q}_\text{max} = 250$~mm/s$^2$ and $1\,000$~mrad/s$^2$ for the linear and rotary axes, respectively. 
The resulting feedrate profiles are shown in \autoref{fig:CompareFeed}, yielding finishing times $t_f$ of $20.24$~s, $16.54$~s and $16.86$~s for the CNC kernel, the one-shot feedrate maximization solution, and the proposed LexLP+SeWin solution, respectively.

Notably, despite the automatic feedrate smoothing achieved by the lexicographic optimization (cf. \autoref{fig:LexSmooth}) and the real-time capable calculation enabled by the sequential windowing strategy, the proposed LexLP+SeWin method increases the finishing time by only $1$~\% compared to the shortest achievable finishing time obtained from the one-shot feedrate maximization problem. 
In contrast, the CNC kernel compromises finishing time by $22$~\% to guarantee real-time execution and computational robustness, as extensively demonstrated in industrial practice. This behavior reflects an inherent trade-off of profile-based feedrate planning, in which the feedrate is scheduled in the workpiece coordinate system, while the axis-level dynamic limits defined in the machine coordinate system can only be enforced approximately or through iterative procedures. 

\begin{figure}[htb]
  \centering
  \includegraphics[width=0.475\textwidth]{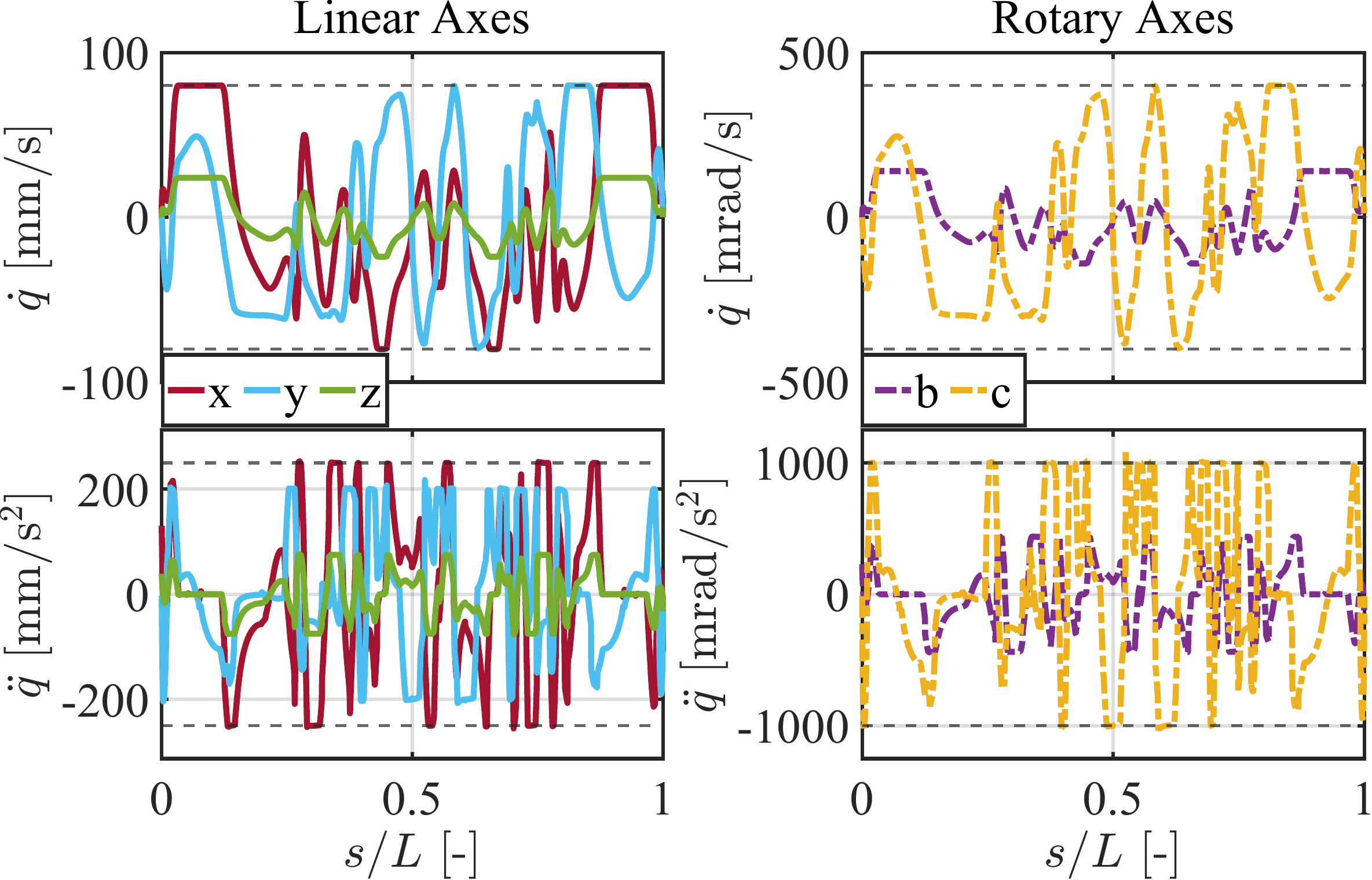}
  \caption{Axis setpoints generated by the proposed method.} 
  \label{fig:AxesTraj}
\end{figure}

\begin{figure}[htb]
  \centering
  \includegraphics[width=0.475\textwidth]{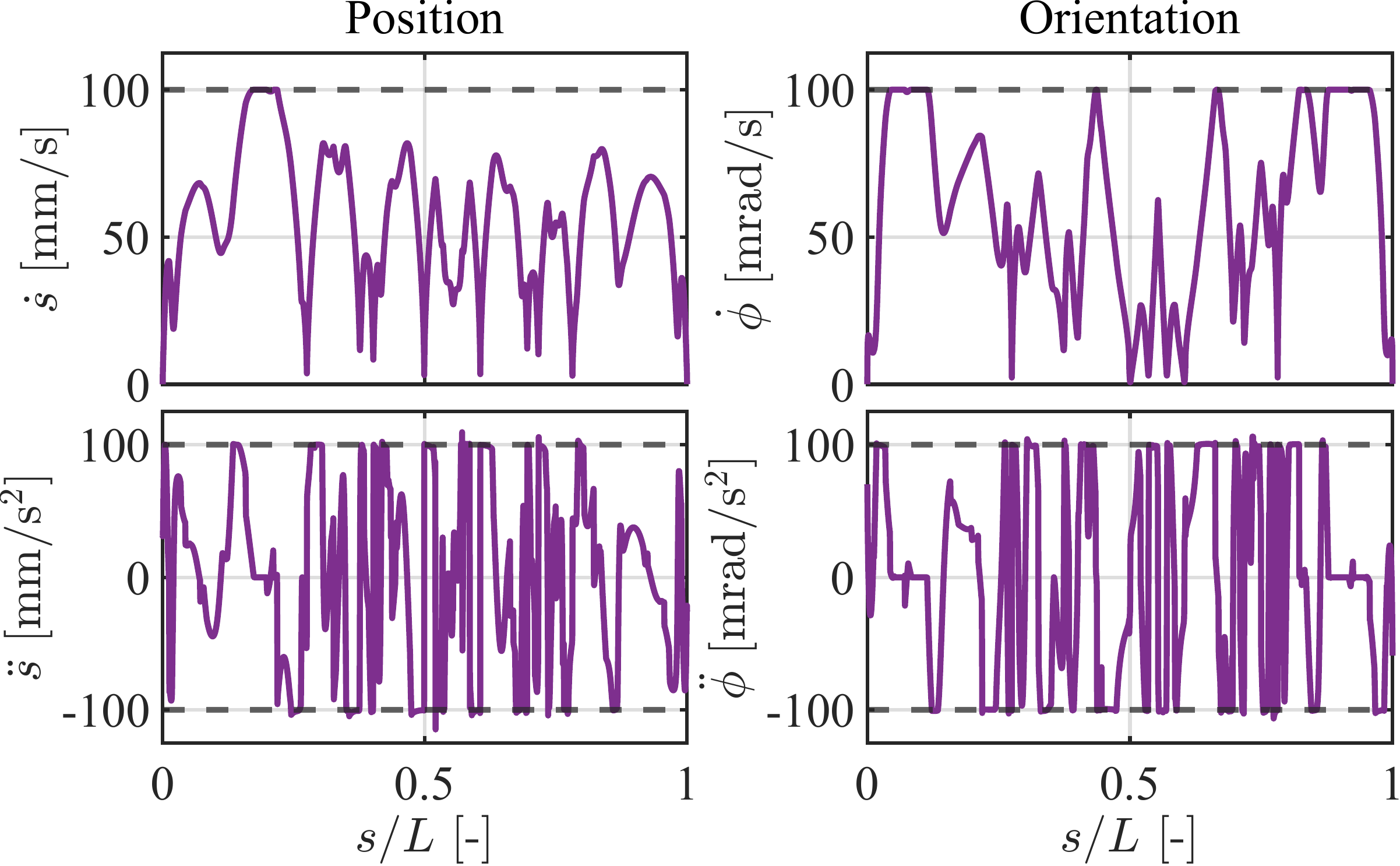}
  \caption{Cartesian motion profiles generated by the proposed method.} 
  \label{fig:TcpTraj}
\end{figure}

\begin{figure}[htb]
  \centering
  \includegraphics[width=0.485\textwidth]{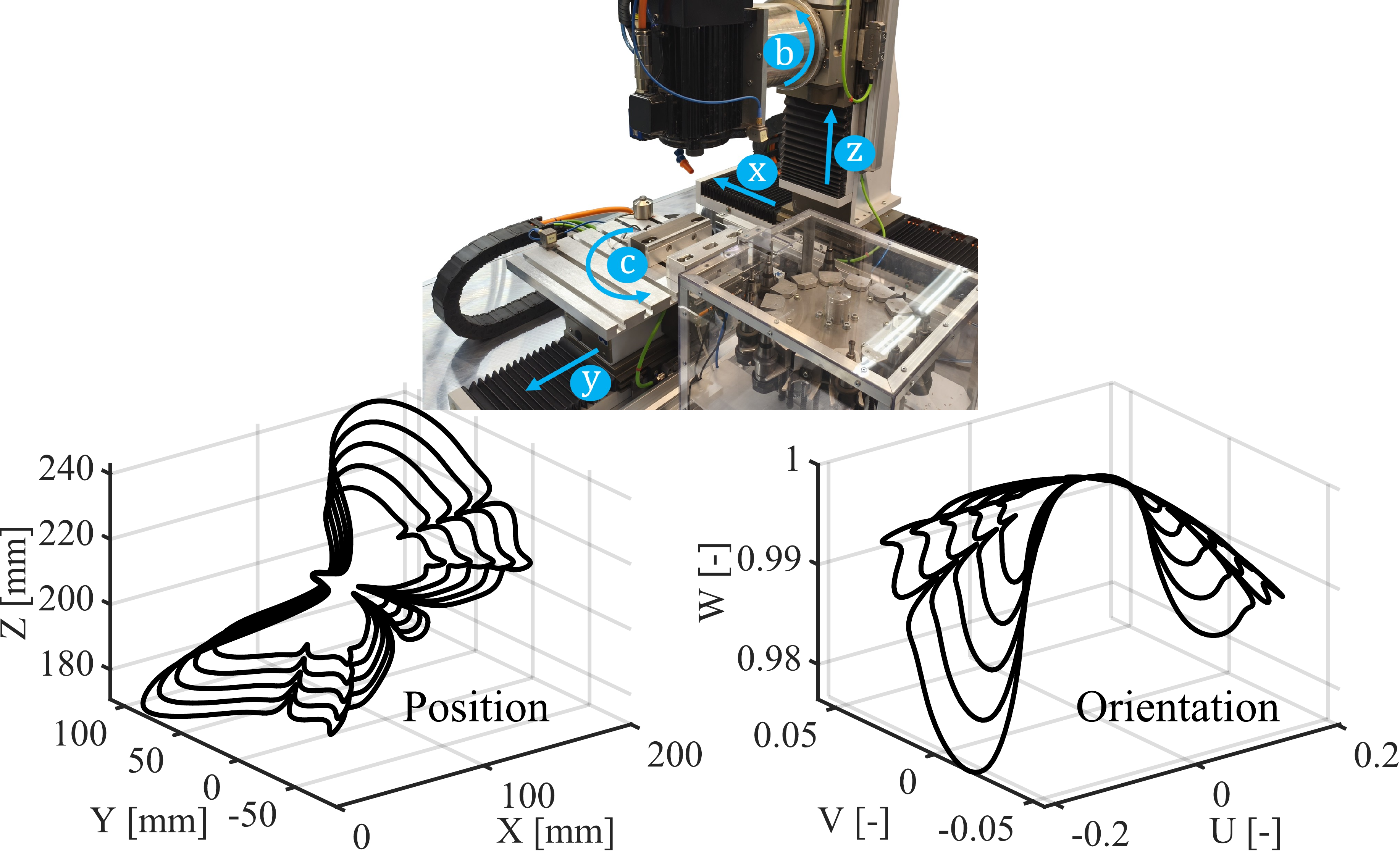}
  \caption{Experimental test bench and test toolpath.} 
  \label{fig:setup}
\end{figure}

To further investigate the constraint feasibility of the presented LexLP optimization method, the interpolated axis-level setpoints in the MCS and the Cartesian feedrate profiles in the WCS are shown in \autoref{fig:AxesTraj} and \autoref{fig:TcpTraj}, respectively. Additional constraints on Cartesian position and orientation are simultaneously imposed. Only minor constraint violations are observed due to fine interpolation to $1$~kHz. Notably, the proposed method does not rely on additional arc-length or arc-radian parameterization to handle Cartesian constraints (cf. \eqref{eq:constrToolPos} and \eqref{eq:constrToolOri}). 
Instead, a unified toolpath parameterization scheme is adopted for both position and orientation in \eqref{eq:path}, ensuring that Cartesian translational and rotational constraints are inherently synchronized, eliminating the need for extra motion synchronization. This further simplifies the geometric preprocessing stage. 

\begin{figure}[htb]
  \centering
  \includegraphics[width=0.485\textwidth]{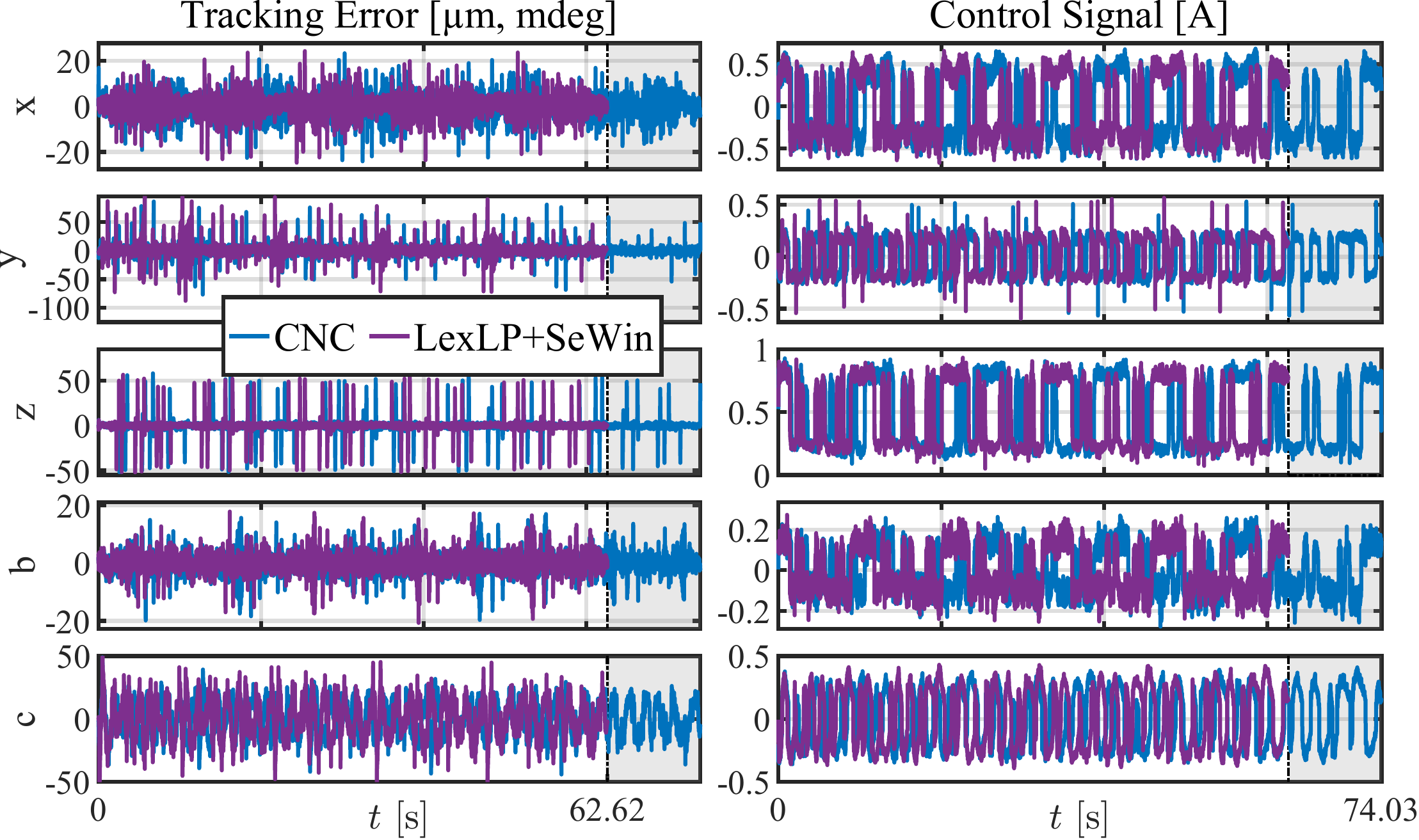}
  \caption{Experimental comparison of tracking error and control signal.} 
  \label{fig:expResult}
\end{figure}

The proposed method was also validated in an experimental test case. A five-axis freeform trajectory with a total arc length of $3766$~mm and arc radian of $232^\circ$ is used, as shown in \autoref{fig:setup}. 
In the LexLP implementation, a total of $N=13\,500$ checkpoints are used, corresponding to a discretization resolution of approximately 0.28 mm. The CNC kernel is also used in the tracking experiments for comparison. The measured experimental tracking errors and control signals for each axis are shown in \autoref{fig:expResult}. 
Despite a reduction in finishing time by $15$~\% compared to the CNC kernel, the proposed LexLP+SeWin method still makes the resulting tracking error at a very similar level. The control signals generated by the LexLP method are slightly larger than those of the CNC kernel, particularly during transient phases where the direction of motor motion changes. 
This behavior results from the less conservative handling of axis limits at the acceleration level, in contrast to the profile-based strategy adopted by the CNC kernel. In addition, this experiment verified the efficiency of the LexLP+SeWin method in handling long toolpaths with a large number of constraint checkpoints on an industrial five-axis machining setup.

\section{Conclusions}
The proposed lexicographic optimization framework achieves an adaptive balance between finishing time reduction and feedrate smoothing without requiring parameter tuning. By exploiting problem sparsity and combining it with a sequential windowing strategy, the method enables real-time capable calculation with verified efficiency and robustness under single-core execution. Its performance is demonstrated on legacy hardware for toolpaths with up to one million constraint checkpoints, confirming linear scalability with respect to toolpath length. Future research will focus on extending the framework to parallel kinematics, which admit analytical expressions for inverse transformation and allow direct formulations in Cartesian space. 

\section*{Acknowlegdements}
This research work was supported by the German Research Foundation (DFG) under project number 557218129. We greatly appreciate the fruitful discussion about kinematic transformation with Dr. Christoph Hinze and Dr. Friedemann Groh.

\bibliographystyle{IEEEtran}
\bibliography{BibSource}

\end{document}